\documentclass[
 reprint,
 amsmath,amssymb,
 superscriptaddress,
 aps,floatfix,
]{revtex4-2}

\usepackage[utf8]{inputenc}
\usepackage{float}
\usepackage[caption=false]{subfig}     
\usepackage{natbib}
\usepackage{graphicx}
\usepackage{xcolor}
\usepackage{booktabs}
\usepackage{multirow}
\usepackage{braket}
\usepackage{hyperref}
\usepackage{tikz}
\usepackage{capt-of}
\usepackage{footmisc}
\usepackage{svg}
\usepackage{mathtools}
\usepackage{url}
\UseRawInputEncoding

\begin{document}

\title{Pairwise-parallel entangling gates on orthogonal modes in a trapped-ion chain}

\author{Yingyue Zhu}
\affiliation{Joint Quantum Institute and Department of Physics, University of Maryland, College Park, MD 20740, USA }
\author{Alaina M. Green}
\affiliation{Joint Quantum Institute and Department of Physics, University of Maryland, College Park, MD 20740, USA }
\author{Nhung H. Nguyen}
\affiliation{Joint Quantum Institute and Department of Physics, University of Maryland, College Park, MD 20740, USA }
\author{C. Huerta Alderete}
\affiliation{Joint Quantum Institute and Department of Physics, University of Maryland, College Park, MD 20740, USA }
\author{Elijah Mossman}
\affiliation{Joint Quantum Institute and Department of Physics, University of Maryland, College Park, MD 20740, USA }
\author{Norbert M. Linke}
\affiliation{Joint Quantum Institute and Department of Physics, University of Maryland, College Park, MD 20740, USA }
\affiliation{Duke Quantum Center and Department of Physics, Duke  University, Durham, NC 27708, USA}
\date{\today}

\begin{abstract}
Parallel operations are important for both near-term quantum computers and larger-scale fault-tolerant machines because they reduce execution time and qubit idling. We propose and implement a pairwise-parallel gate scheme on a trapped-ion quantum computer. The gates are driven simultaneously on different sets of orthogonal motional modes of a trapped-ion chain. We demonstrate the utility of this scheme by creating a GHZ state in one step using parallel gates with one overlapping qubit. We also show its advantage for circuits by implementing a digital quantum simulation of the dynamics of an interacting spin system, the transverse-field Ising model. This method effectively extends the available gate depth by up to two times with no overhead apart from additional initial cooling when no overlapping qubit is involved. This is because using a set of extra modes as additional quantum degrees of freedom is nearly equivalent to halving the trap heating rate, doubling the laser and qubit coherence time, and extending the controller memory depth by up to a factor of two. This scheme can be easily applied to different trapped-ion qubits and gate schemes, broadly enhancing the capabilities of trapped-ion quantum computers.
\end{abstract}
\maketitle

\section{Introduction}
In order to create quantum computers that can tackle problems of practical value, it is essential to make optimal use of the quantum resources available in a given technology. Realizing quantum gates in parallel is important for scaling quantum devices since it speeds up the implementation of quantum circuits and allows for more operations to be executed within the coherence time. 

Trapped ions are one of the most advanced quantum computing platforms due to their high-fidelity quantum operations, all-to-all connectivity, and long coherence times \cite{coherencetime_KKim,Debnath2016}. Parallel entangling gates have been reported in trapped-ion experiments \cite{CFiggatt_PG,EASE,EASE_circuit,GG_Kim}. However, these schemes require extra classical overhead to calculate specific gate solutions for each unique combination of gates. They also need extra optical power or longer gate times, and, as a result, come with a loss in fidelity compared to sequential gates.

In harmonically trapped ion chains, entanglement generation between the spin qubits is mediated by the collective motion of the ion crystal \cite{Bruzewicz-2019}, which can be excited with laser beams \cite{MS1998June,Leibfried_2003}, microwaves \cite{Ospelkaus-2008}, or radio frequency waves \cite{laserfree-static}. 
In all trapped-ion entangling gate schemes realized so far, parallel or sequential, only one set of motional modes is considered, corresponding to the direction of strongest overlap between the driving field and a principle trap axis. However, the ion chain has multiple sets of independent motional modes along orthogonal directions in space. The other sets of motional modes are usually ignored, or decoupled by a rotation of the trap principle axes.  While used in a few analog quantum simulations \cite{Gorman,Cinthia-analog} and considered in cost estimation for error correction \cite{Trout_2018}, this resource remains untapped for entangling gates and digital quantum circuits. In this work, we use an additional set of modes to implement two entangling gates on arbitrary pairs of ions in parallel, including cases where there is one overlapping ion between the two pairs, with Raman laser beams \cite{Debnath2016}. 
Since the two sets of modes are orthogonal, there is no need to calculate unique parallel gate pulses, as existing laser pulse designs can simply be applied simultaneously. For addressing beam arrays, no extra laser power is required when there is no overlapping ion since idle energy in the laser beams is used. For overlapping pairs, this scheme may require twice the laser power of a single entangling gate on the shared ion, but typically less than that. The idea is directly applicable to any addressed entangling gate scheme that uses the ion motion \cite{MS1998June,Benhelm-2008,CZ_1995,EASE,EASE_circuit,GG_Kim,CFiggatt_PG,Katz22,CZ-CnNOT,Shapira23,Katz22,zzGate,Chao-2023,Leibfried_2003,Baldwin-2021}. It can also be extended to the third motional direction.

\section{parallel gate scheme}
\begin{figure}
    \centering
    \includegraphics[scale=0.5]{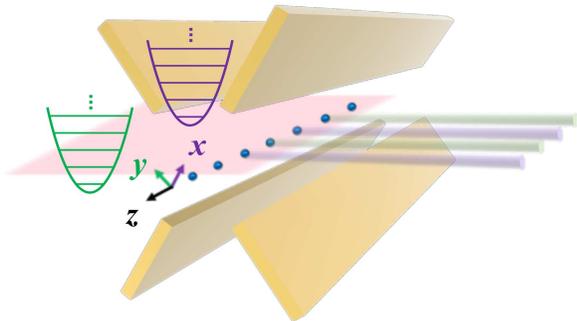}
    \caption{Experimental setup for parallel MS gates on orthogonal motional modes. A chain of ions is trapped in a linear Paul trap indicated by the four trap electrodes. The right and left edge ions are used as endcaps, and the middle five ions are used as qubits. $x$, $y$, and $z$ are the three principal axes of the trap. The harmonic oscillator modes along the $x$- and $y$-axis are shown in purple and green, respectively. Transitions are driven by pair of Raman beams consisting of a global beam (pink) and a counter-propagating addressing array. The colors of the individual beams indicate different detunings for exciting motional modes along different principal axes. This figure shows an MS gate on the first and third qubit from the left via the $x$ motional modes driven in parallel with an MS gate on the second and fourth ion via the $y$ motional modes. }
    \label{fig:trap}
\end{figure}

In a linear Paul trap, the ions form a chain.  In the radial directions, i.e. the plane perpendicular to the chain, the ions are trapped in a pseudo-potential well that is approximately harmonic along the two radial principal axes, $x$ and $y$. The ions experience a weaker harmonic confinement along the chain, in the axial $z$ direction \cite{James_1997}.  The $N$ ions in the chain act as a set of coupled harmonic oscillators along each principal axis as a result of the confining potential and the Coulomb repulsion between the ions. This gives rise to $N$ normal motional modes along each of the $x$, $y$, and $z$ directions. Normal modes on different principal axes are independent \cite{ionmotion1,James_1997}. 

Figure \ref{fig:trap} shows the concept for running pairwise parallel gates. In the depicted setup, the ion chain is illuminated by an array of individual addressing beams and one counterpropogating global beam for driving coherent operations, and the beam directions overlap with the two radial principle axes. We use the M\o{}lmer-S\o{}rensen (MS) scheme as our two-qubit gates. Using different frequencies, one pair of beams drives an MS gate on the $x$-modes (shown in green), and another pair simultaneously does the same on another pair of ions on the $y$-modes (shown in purple). This protocol can be applied to addressed trapped-ion chains in any trap geometry, not limited to the one shown in Figure \ref{fig:trap}, provided that the driving field has non-zero projections along multiple principal axes. Other gate schemes can also be used, as noted in the introduction. 

Here we describe our experiment and protocol as an example. We trap $^{171}$Yb$^{+}$ ions in a linear Paul trap. The qubit states $|0\rangle$ and $|1\rangle$ are encoded in the two hyperfine ground states $\ket{F=0,m_F=0}$ and $\ket{F=1,m_F=0}$ in the $^{2}\text{S}_{1/2}$ manifold. 
The laser beams used for coherent operations are derived from a mode-locked laser and drive a Raman transition between the two qubit states \cite{SDebnath_2016}. The beat-notes between the two beams are at $\omega_{0}\pm\mu$, which are closely detuned from blue and red motional sidebands \cite{Debnath2016}, where $\omega_{0}$ is the carrier frequency. In order to disentangle the modes from the qubits at the end of the gate, modulation schemes for the amplitude, phase, frequency of the laser pulse or a combination of these are employed, leaving only the qubit states of the ion pair in entanglement \cite{AM1,AM2_Roos_2008,FM1,PM1,AMFM}. We use amplitude modulation \cite{Choi2014} for this demonstration, but any of the above schemes can equally be used. The unitary representing an MS gate on the qubit pair $\{p,q\}$ is
 $U=\exp\left( i\chi_{pq}\sigma_{x}^{p}\sigma_{x}^{q} \right)$, where $\sigma_{x}^{p}$ is the Pauli-$X$ operator acting on qubit $p$, and $\chi_{pq}$ is the gate angle which can be varied arbitrarily by applying a scale factor to the amplitude of laser pulse.
 
While one MS gate on the pair $\{p,q\}$ is driven on the $x$ motional modes, we simultaneously drive another MS gate via the $y$ modes on the qubit pair $\{m,n\}$. This simultaneous gate can be implemented in the same way as described above with the beat-notes of the counter-propagating Raman beams tuned to $\omega_{0}\pm\mu'$, which are closely detuned from the motional side-bands of the $y$ modes. We now show that these simultaneous operations do not introduce cross-couplings. In the interaction picture and the Lamb-Dicke regime, neglecting the off-resonant carrier term, the Hamiltonian describing the interactions present during the parallel gate (PG) is 
\begin{equation}\label{H_PG}
    H_{\text{PG}}(t)=H_{x}(t)+H_{y}(t),
\end{equation}
where
\begin{equation*}
\begin{split}
      &H_{x}(t)=\!\!\!\!\sum_{i\in\{p,q\}}\sum_{k=1}^{N}\Omega_{i}(t)\eta_{k}^{i}\text{cos}(\mu t -\phi_{i})(a_{k} e^{-i\omega_{k}t}+h.c.)\sigma_{x}^{i} \\
      &\text{and}\\
    &H_{y}(t)=\!\!\!\!\!\!\sum_{j\in\{m,n\}}\!\sum_{k'=1}^{N}\Omega_{j}(t)\eta_{k'}^{j}\text{cos}(\mu' t -\phi_{j})(b_{k'} e^{-i\omega_{k'}t}+h.c.)\sigma_{x}^{j}.
\end{split}
\end{equation*}
$N$ is the total number of ions in the chain. $\omega_{k}$($\omega_{k'}$) is the frequency of the motional mode $k$($k'$) in the $x$($y$)-direction and $a_{k}^{\dagger}$ and $a_{k}$($b_{k'}^{\dagger}$ and $b_{k'}$) are the creation and annihilation operator for mode $k$($k'$) in the $x$($y$)-direction. $\Omega_{i}$($\Omega_{j}$) is the Rabi frequency of qubit $i$($j$). $\eta_{k}^{i}$($\eta_{k'}^{j}$) is the Lamb-Dicke parameter coupling qubit $i$($j$) to mode $k$($k'$) in $x$($y$). $\phi_{i}$($\phi_{j}$) is determined by the laser phase.

In order to calculate the evolution unitary, we exponentiate Eq.~\eqref{H_PG} using the Magnus expansion. Looking at the first two terms in the Magnus series, we have
\begin{multline}\label{magnus}
  \!\!\!\!\!\!U_{\text{PG}}(\tau) = e^{-i\int_{0}^{\tau}dtH_{\text{PG}}(t)
  -\frac{1}{2}\int_{0}^{\tau}dt_{2}\int_{0}^{dt_{2}}dt_{1}[H_{\text{PG}}(t_{2}),H_{\text{PG}}(t_{1})]},
\end{multline}

\noindent where $\tau$ is the gate time. Expanding the commutator yields
 \begin{equation}\label{commutator}
    \begin{split}
     [H_{\text{PG}}(t_{2}),H_{\text{PG}}(t_{1})]=&
    [H_{x}(t_{2}),H_{x}(t_{1})]+[H_{y}(t_{2}),H_{y}(t_{1})]\\
    +&[H_{x}(t_{2}),H_{y}(t_{1})]+[H_{y}(t_{2}),H_{x}(t_{1})].
     \end{split}
 \end{equation}
 Note that, $\sigma^{i}_{x}$ commutes with $\sigma^{j}_{x}$, and since the $x$ and $y$ modes are orthogonal $a_{k}$ and $a_{k}^{\dagger}$ commute with $b_{k'}$ and $b_{k'}^{\dagger}$. As a result, $[H_{x}(t_1)$,$H_{y}(t_2)]=0$ at any time as well as the last two terms in Eq.~\eqref{commutator}, which are the cross-coupling terms between $x$ and $y$. We emphasize that the scheme also applies to cases where the two qubit pairs share an ion since $[\sigma^{i}_{x},\sigma^{j}_{x}]=0$ includes the case of $i=j$. Either remaining term in Eq.~\eqref{commutator} contains motion in the $x$- direction only or the $y$-direction only.
Same as the case of one MS gate, the dependence on motion eventually vanishes in Eq.~\eqref{commutator} due to the commutation relations between the motional operators. Therefore, the higher-order terms not shown in the Magnus expansion in Eq.~\eqref{magnus} are also zero. We can rewrite Eq.~\eqref{magnus} as

\begin{multline} \label{eq:U_PG}
    U_{\text{PG}}(\tau)=\textrm{exp}\left( \sum_{i\in\{p,q\}}(\emph{G}_{i}\sigma_{x}^{i})+i\chi_{pq}\sigma_{x}^{p}\sigma_{x}^{q}\right.\\
   \left. +\sum_{j\in\{m,n\}}(G^{\prime}_{j}\sigma_{x}^{j})+i\chi_{mn}\sigma_{x}^{m}\sigma_{x}^{n}\right),
\end{multline}
where $G_i$ is defined as
\begin{equation}\label{eq:displacement}
    G_{i}= \sum_{k=1}^{N}\alpha_{i,k}(\tau)a_{k}^{\dagger}+\alpha^{\star}_{i,k}(\tau)a_{k}
\end{equation}
\begin{equation}
        \alpha_{i,k}(\tau)= -\int^{\tau}_{0}\eta^{i}_{k}\Omega_{i}(t)\cos(\mu t-\phi_{i}^{m})e^{i\omega_{k}t}dt.
\end{equation}

The expression for $G^{\prime}_{j}$ is similar to Eq.~\eqref{eq:displacement} except that the sum over $x$ modes is replaced by a sum over $y$ modes and $\mu$ is replaced by $\mu'$. Since the first two terms in the exponential in Eq.~\eqref{eq:U_PG} only involve $x$ modes while the last two terms only involve $y$ modes, these two parts can be treated separately. 
The Rabi frequencies $\Omega_{i}(t)$($i\in\{p,q\}$) are modulated such that $G_{i}(\tau)=0$. The pulse modulation is done likewise for the qubit pair $\{m,n\}$ to set $G'_{j}(\tau)=0$ ($j\in\{m,n\}$).
The gate angles $\chi_{pq}$ and $\chi_{mn}$ can be set independently. Finally, the resulting unitary Eq.~\eqref{U_PG_final} describes parallel MS gates on $\{q,p\}$ and $\{m,n\}$ with angle $\chi_{pq}$ and $\chi_{mn}$, respectively:


\begin{equation}\label{U_PG_final}
    U_{\text{PG}}(\tau)=\textrm{exp}(i\chi_{pq}\sigma_{x}^{p}\sigma_{x}^{q} +i\chi_{mn}\sigma_{x}^{m}\sigma_{x}^{n})
\end{equation}.

\begin{figure}
    \centering
    \includegraphics[width=\linewidth]{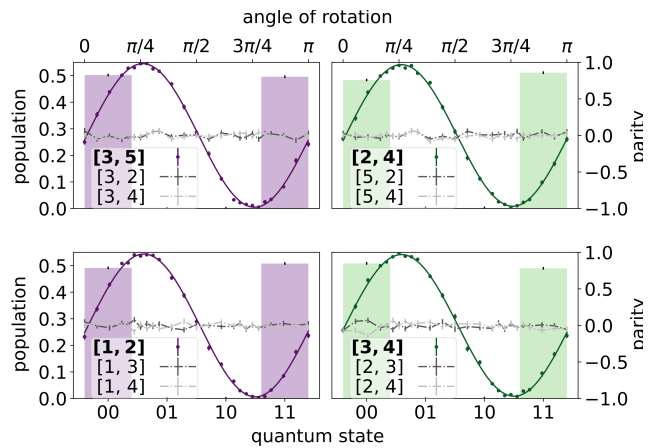}
    \caption{Fidelity measurements for two representative pairs of parallel MS gates on non-overlapping pairs of qubits in a seven-ion chain. The populations in the computational basis (bottom abscissa, left ordinate) and the parity scan (bottom abscissa, right ordinate) are plotted for each pair. The top row shows pair \{3,5\}, in purple, entangled via the $x$ modes, and pair \{2,4\}, in green, entangled via the $y$ modes. Purple and green solid curves are fitted to the parity scan data. The measured fidelities are $F_{35}=99.1(4)\%$ and $F_{24}=98.2(4)\%$. Parity measurements of the cross-talk pairs plotted in grey scale in the background show no oscillation. The bottom row of the figure shows pair \{1,2\} ($x$ modes) and \{3,4\} ($y$ modes), where $F_{12}=98.8(5)\%$ and $F_{34}=98.4(4)\%$. All uncertainties are statistical.}
    \label{fig:nonoverlapping_fidelity}
\end{figure}

\section{Experimental results}
Using a chain of seven ions, we experimentally demonstrate parallel MS gates on different qubit pairs, including the case where one qubit is shared between the two pairs. We then show their benefit for long circuits by running a digital quantum simulation of a paradigmatic spin model, the transverse field Ising model.

\subsection{Parallel gate fidelity}

\begin{figure}
    \centering
    \includegraphics[width=\linewidth]{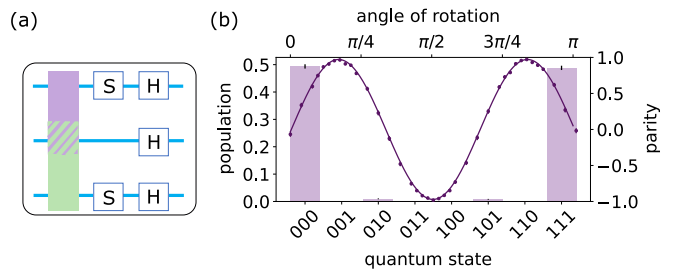}
    \caption{(a) Circuit for preparing a three-qubit GHZ state with parallel gates. All qubits are initiated in the $|0\rangle$ state. The first gate is a pair of parallel MS gates overlapping on the middle qubit. $H$ is the Hadamard gate. The $S$ gate is defined as $S=e^{-i\sigma_{z}\frac{\pi}{4}}$. (b) Parity contrast (top abscissa, right ordinate) and populations (bottom abscissa, left ordinate) of an experimentally prepared GHZ state on a seven-ion chain. The estimated fidelity is $F_{\text{GHZ}}=97.6(3)\%$. Error bars are statistical.}
    \label{fig:GHZ}
\end{figure}

\begin{figure*}
    \centering
    \includegraphics[width=\textwidth]{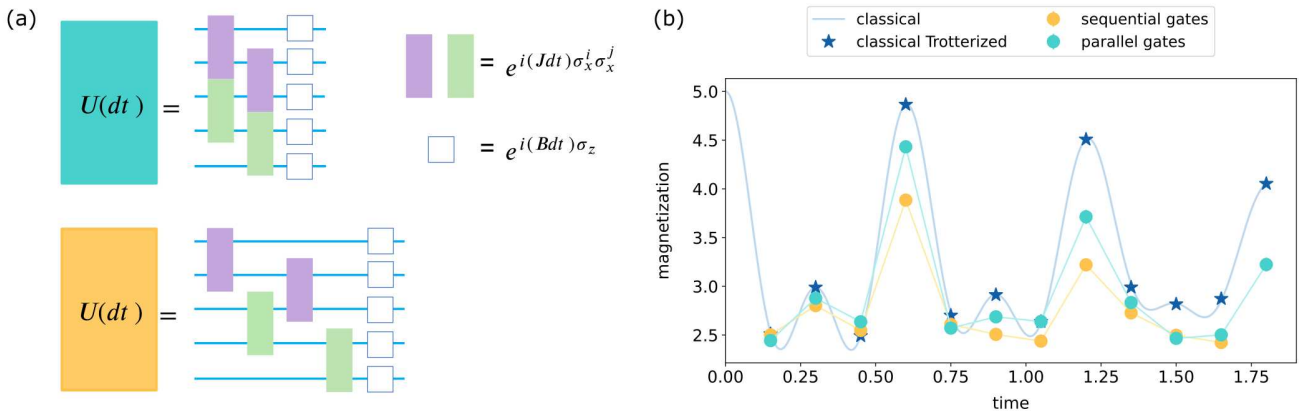}
    \caption{(a) TFIM circuits of one Trotter step with parallel gates (top) and sequential gates (bottom), using the $x$ modes (purple) and $y$ modes (green). (b) TFIM experimental results, where $\frac{B}{J}=0.096$. The total magnetization $m$ of the spin chain in the $z$ direction is evaluated at different evolution times. The evolution with parallel MS gates shows a much higher contrast than the evolution with sequential gates. The error bars on the experimental points are statistical and are smaller than the marker.}
    \label{fig:TFIM}
\end{figure*}

For two parallel MS gates of angle $\pi/4$ implemented on two non-overlapping pairs of qubits, we estimate the fidelity of each MS gate by measuring the even-parity state populations of the output state and the parity contrast \cite{fidelity1,C_Ballance}. The parity contrast is extracted by appending single-qubit $\pi/2$ rotations to the MS gate and varying their phases. The results are shown in 
Fig.~\ref{fig:nonoverlapping_fidelity}. We find that the performance of our parallel gates is overall very similar to that of our sequential gates (see Table~\ref{SequentialFidelity}), indicating that no detectable error, such as motional cross-talk between the two sets of radial modes due to trap non-linearities, is added by the parallel operation. Yet, the reduction in execution time will improve the fidelity of a complete circuit implementation with parallel gates. We also estimate the degree of entanglement of the cross-pairs. This could occur as a result of optical cross-talk or motional cross-talk. Parity measurements plotted in gray scale in Fig.~\ref{fig:nonoverlapping_fidelity} show no oscillation within errors, which indicates no unwanted entanglement between the cross-talk pairs.  

      
\begin{table}[h]
    \begin{tabular}{p{1cm} p{1.5cm} p{1.5cm}}
    \hline
    Pair&\hspace{0.25cm}PG&Sequential\\
    \hline
    XX35    &$99.1(4)\%$&   $99.1(3)\%$ \\   
    XX24    &$98.2(4)\%$&   $99.2(4)\%$ \\   
    XX12    &$98.8(5)\%$&   $97.5(4)\%$\\
    XX34    &$98.4(4)\%$&   $99.2(4)\%$\\      
    \hline
    \hline
    \end{tabular}
     \caption{\label{SequentialFidelity}The table compares parallel gate (PG) fidelities to sequential gate fidelities.}
\end{table}

\subsection{Three-qubit entanglement}

We use our scheme to entangle three ions in one step, generating a three-qubit GHZ state. The circuit is shown in Fig. ~\ref{fig:GHZ}(a). It consists of a three-qubit entangling gate followed by $\pi/2$ phases gates and Hadamard gates.

Fig.~\ref{fig:GHZ}(b) shows the populations and parity contrast of the GHZ state prepared with parallel MS gates on qubits $\{3,5\}$ and $\{2,5\}$. We estimate the fidelity using parity contrast and even-parity state populations again to find $F_{\text{GHZ}}=97.6(3)\%$, which is as good as the three-qubit GHZ of $97.1(1)\%$ fidelity prepared on the same system with sequential MS gates and improves upon the three-qubit GHZ state prepared with global parallel gates reported on another trapped ion experiment \cite{GG_Kim}. 
Since the laser pulses for both MS gates are summed up on the overlapping ion, in the worst case scenario, it will require the sum of the power of the two gates to implement them in parallel. However, it can often be done with less power because the gate pulse segments with the highest amount of power in the two gate pulse shapes do not always overlap. 

\subsection{Transverse field Ising model with parallel gates}\label{TFIM}
Quantum spin models can describe a wide range of quantum many-body dynamics and are suitable for implementation on different hardware platforms \cite{Spin_model_Kim_2011,SpinModel_Schauss_2018}. They can be used to study a variety of classically intractable problems in condensed matter physics, such as quantum magnetism \cite{qMagnetism}, spin glasses \cite{qSpinGlass}, and others \cite{RevModPhys.93.025001}. We demonstrate the experimental advantage of our parallel gate scheme over sequential gates by digitally simulating the dynamics of 1D Transverse Field Ising Model (TFIM) of five spins with nearest-neighbor interactions. The Hamiltonian can be written as
\begin{equation}
    H=-J\sum_{i=1}^{4}\sigma_{x}^{i}\sigma_{x}^{i+1}-B\sum_{i=1}^{5}\sigma_{z}^{i},
\end{equation}
where $J$ and $B$ characterize the strength of interaction between the neighboring spins and the magnitude of the external field, respectively. The spin $|\uparrow\rangle$ ($|\downarrow\rangle$) state is mapped to the qubit $|+\rangle$($|-\rangle$) state. All qubits are initialized in the $|0\rangle=\frac{1}{\sqrt{2}}(|\uparrow\rangle+|\downarrow\rangle)$ state. The time-evolution unitary is Trotterized and decomposed into gates \cite{Lloyd-1996,Suzuki-1991}. Circuits representing one Trotter step are shown in Fig.~\ref{fig:TFIM} (a). Both circuits consist of MS gates, applied in sequence or in parallel, and single-qubit z-rotations. After $N$ Trotter steps, the qubits are measured in the $z$ basis, and the total magnetization along $z$, $m=\sum_{i=1}^{5}\sigma_{z}^{i}$, is calculated. Since the single-qubit $z$-rotations are done as instantaneous classical phase advances on the controller, the circuit execution time with sequential MS gates is twice as long as that with parallel MS gates. 

 The results are plotted in Fig.~\ref{fig:TFIM} (b). In the experiment, the sequential gates are also done on both $x$ and $y$ motional modes, consistent with the parallel gates, to ensure a fair comparison. The same MS gate pulse is used for each qubit pair in both the parallel implementation and the sequential implementation. For points of high-magnetization, for which the observable is most sensitive to errors, we see a reduction in error by up to a factor of two in the parallel gate implementation as a consequence of halving the circuit run time using parallel gates. This demonstrates that there is an advantage in using parallel gates, even for shorter circuits. 

\section{Conclusion and outlook}
While employing these pairwise-parallel gates in an ion trap quantum computer does not reduce algorithm complexity, it can provide a significant fidelity boost to any circuit implementations and should be used whenever possible. The only cost is a fixed overhead for ground-state cooling the additional set of modes, which is already cooled to below the Doppler limit by the high-intensity laser sideband cooling of the first set of modes. There is no calibration overhead in the parallel implementation. In fact, the required number of calibrations is usually reduced as gates can be calibrated in parallel. In the worst case, where a gate for each qubit pair needs to be calibrated with both sets of modes, the number of calibrations is the same as in the sequential case. 

This parallel gate is easily transferable to other linear trap geometries. It can be generalized to any MS gate pulse modulation and addressing schemes, provided that the controllers can target multiple ions in space and multiple modes in frequency. It could also be extended to the axial direction and be combined with other parallel gate schemes that use motional modes only along one principal axis \cite{EASE,GG_Kim,CFiggatt_PG} to further increase the maximum number of simultaneous two-qubit gates. Finally, other entangling gate mechanisms, such as light shift gates \cite{Leibfried_2003,Baldwin-2021}, those driven by a synthetic $\sigma_z$ spin-dependent force \cite{zzGate}, the multi-qubit gate based on the Cirac-Zoller scheme with ancillary motional states \cite{Chao-2023}, or gates based on mode squeezing via second-sideband driving \cite{Katz22, Shapira23} can be pairwise-parallelized in the same way.

\section{Acknowledgements}

This material is based upon work supported by the U.S. Department of Energy, Office of Science, National Quantum Information Science Research Centers, Quantum Systems Accelerator. Additional support is acknowledged from the Office of Naval Research (N00014-20-1-2695) and the National Science Foundation (QLCI grant OMA-2120757). We thank Liam Jeanette for assistance with the data collection and Kenneth R. Brown for helpful discussions.  

\bibliographystyle{ieeetr}
\bibliography{reference}
\end{document}